\documentclass[preprint,12pt]{elsarticle}
\usepackage{amsmath}
\usepackage[cmtip,arrow]{xy}
\usepackage{pb-diagram,pb-xy}
\usepackage{slashed}
\usepackage{amssymb,amsopn}
\def \be {\begin{equation}}
\def \ee {\end{equation}}
\def \bea {\begin{eqnarray}}
\def \eea {\end{eqnarray}}
\def \sla {\slashed}
\usepackage{graphicx}

\journal{Physics Letters B}

\begin{document}

\begin{frontmatter}

\title{Higher-order Lorentz-invariance violation, quantum gravity and fine-tuning }

%\author{C. M. Reyes, L. F. Urrutia and J. D. Vergara}

%\address{Instituto de Ciencias Nucleares, Universidad Nacional Aut{\'o}noma de M{\'e}%
%xico, \\
%A. Postal 70-543, 04510 M{\'e}xico D.F., M{\'e}xico}

%\author{ Carlos M. Reyes$^{1}$, Sebastian Ossandon$^{2}$ and  Camilo Reyes$^{3}$}
%\email[Electronic mail: ]{creyes@ubiobio.cl}
%\address{$^{1}$ Departamento de Ciencias B{\'a}sicas, Universidad 
%del B{\'i}o B{\'i}o, Casilla 447, Chill\'an, Chile}
%\address{$^{2}$ Instituto de Matem\'aticas, Pontificia Universidad 
%Cat\'olica de Valpara\'iso, Casilla 4059, Valpara\'iso, Chile}
%\address{$^{3}$ Departamento
%de Ciencias Fisicas, Facultad de Ciencias Exactas, Universidad Andres Bello, Republica 220, 
%Santiago, Chile}

\author{Carlos M. Reyes}
%\ead[url]{creyes@ubiobio.cl}
\address{Departamento de Ciencias B{\'a}sicas, Universidad 
del B{\'i}o B{\'i}o, Casilla 447, Chill\'an, Chile}
\author{Sebastian Ossandon}
\address{Instituto de Matem\'aticas, Pontificia Universidad 
Cat\'olica de Valpara\'iso, Casilla 4059, Valpara\'iso, Chile}

\author{Camilo Reyes}
\address{Departamento
de Ciencias Fisicas, Facultad de Ciencias Exactas, Universidad Andres Bello, Republica 220, 
Santiago, Chile}

\begin{abstract}
The issue of Lorentz fine-tuning in effective theories containing higher-order operators is studied. 
To this end, we focus on the Myers-Pospelov extension of QED with
dimension-five operators in the photon sector and standard fermions.
We compute the
fermion self-energy at one-loop order considering its
even and odd $CPT$ contributions. In
the even sector we find small
 radiative corrections to the usual parameters of QED
which also turn to be finite.
In the
odd sector the axial operator is shown to contain unsuppressed effects of Lorentz violation leading to a possible
fine-tuning.
We use dimensional regularization to deal with the divergencies and 
a generic preferred four-vector.
Taking the first steps in the renormalization procedure for Lorentz violating theories
we arrive to 
 acceptable small corrections allowing to set the bound 
 $\xi<6 \times10^{-3}$.
\end{abstract}


\end{frontmatter}

\section{Introduction}
New physics from the Planck scale has been hypothesized to 
show up at low energies as small 
 violations of Lorentz symmetry~\cite{foam}.
 This possibility has been supported by the idea that  spacetime may
 change drastically at high energies
 giving place to some level or discreteness or spacetime foam.
In the language of effective theory the Lorentz symmetry departures are implemented
with   
Planck mass suppressed operators in the 
Lagrangians. 
The effective approach has been shown to be extremely successful in order
to contrast the 
possible Lorentz and CPT symmetry violations 
with experiments. 
A great part of these searches have been given within the framework of 
the standard model extension with 
 several bounds on Lorentz symmetry violation provided~\cite{SME,Table,C-S}.
In general most of the studies on 
Lorentz symmetry violation have been performed
with operators of mass dimension $d\leq 4$,~\cite{C-S-radiative}.
In part because the higher-order 
theories present some problems in their quantization~\cite{unitarity}.
However, in the last years these operators 
 have received more attention
and several bounds have been
put forward~\cite{Antoniadis,Urrutia,Mariz,Casana,petrov}. 
Moreover, a generalization has been constructed
to include non-minimal terms in the effective framework of
the standard model extension~\cite{KM1}.

Many years ago Lee and Wick~\cite{Lee-Wick} and Cutkosky~\cite{CUT} 
studied the unitarity of higher-order theories using the formalism
of indefinite metrics in Hilbert space.
They succeeded to prove that unitarity can be conserved
in some higher-order models
by restricting the space of asymptotic states.
This has stimulated the 
construction of several higher-order models 
beyond the standard model~\cite{Grinstein}.
One example is the Myers and Pospelov model based on
dimension-five operators describing possible 
 effects of quantum gravity 
~\cite{MP,OneloopUnitarity}. 
In the model the Lorentz symmetry violation is characterized
by a
 preferred 
four-vector $n$~\cite{cutoff,causality}. The preferred 
four-vector may be thought to come from a spontaneous symmetry breaking in an
underlying fundamental theory.
One of the original motivations to incorporate such terms
was to produce
cubic modifications in the dispersion relation, although 
an exact calculation yields a more complicated structure usually with the 
gramian of the two vectors $k$
and $n$ involved.
The Myers and Pospelov model has become an important arena 
to study higher-order effects of Lorentz-invariance
 violation~\cite{Urrutia,Crichigno,astro,bret}. 

This work aims to contribute to the discussion 
on the fine-tuning problem due to Lorentz symmetry violation~\cite{fine-tuning},
in particular when higher-order operators are present.
There are different approaches to the subject, for example 
 using the ingredient of discreteness~\cite{discrete} 
or supersymmetry~\cite{Jain}. 
 For renormalizable operators, including 
 higher space derivatives, large Lorentz violations
 can or not appear depending on the model and regularization scheme~\cite{Anselmi}.
 However, higher-order operators are good candidates to produce
strong Lorentz violations via induced lower dimensional operators~\cite{Bolokhov}. 
Some attempts to deal with the fine tuning problem considers modifications
in the tensor contraction with a given
 Feynman diagram~\cite{MP} or just
restrict attention to
higher-order corrections~\cite{nonrenormalizable-op}.
However in both cases the problem comes back at higher-order
 loops~\cite{Perez}. 
Here we analyze higher-order Lorentz violation by
explicitly  computing the radiative corrections 
 in the Myers and Pospelov extension of QED.
We use 
dimensional regularization which eventually preserves unitarity, thus extending
some early treatments~\cite{cutoff,Crichigno}.
%-------------------------------------------------------------------------
\section{The QED extension with dimension-5 operators}
%-------------------------------------------------------------------------
The Myers-Pospelov Lagrangian 
extension of QED with modifications in the
photon sector can be written as~\cite{MP}
\begin{eqnarray}\label{Lag}
\mathcal L&=&\bar \psi(\gamma^{\mu}\partial_{\mu}-m)
\psi-\frac{1}{4}F^{\mu\nu}  F_{\mu\nu}- \frac{\xi}{2m_\text{Pl}} 
n_{\mu}\epsilon^{\mu\nu \lambda \sigma} A_{\nu}(n 
\cdot \partial)^2   F_{\lambda \sigma}\;,
\end{eqnarray}
where $m_\text{Pl}$ is the Planck mass,
$\xi$ a dimensionless coupling parameter 
and $n$ is a four-vector defining a preferred reference frame.
In addition we introduce the gauge fixing Lagrangian term,
$
\mathcal L_{G.F}=-B(x) (n\cdot A)\;,
$
where $B(x)$ is an auxiliary field.

The field equations 
for $A_{\mu}$ and $B$
derived from the Lagrangian $\mathcal L+\mathcal L_{G.F}$ read,
\begin{eqnarray}\label{eqmot}
\partial_{\mu} F^{\mu \nu}+g \epsilon ^{\nu \alpha \lambda \sigma} n_{\alpha} 
 (n \cdot \partial)^2   F_{\lambda \sigma}  &=&B n^{\nu}\;, 
 \\
 n\cdot A&=&0 \label{eqmot2} \;.
\end{eqnarray}
where $g=\frac{\xi}{m_\text{Pl}}$.
Contracting Eq. (\ref{eqmot}) with $\partial_{\nu}$ gives
$
(\partial \cdot n) B=0\;,
$
which allows us to set
$
 \quad B=0\;.
$
In the same way, the contraction of Eq. (\ref{eqmot}) with 
$n_{\nu}$ in momentum space
leads to $k \cdot  A=0\;.$

We can choose the polarization vectors $e_\mu^{(a)}$ with $a=1,2$
to lie
on the orthogonal hyperplane defined by $k$ and $n$~\cite{AA}, satisfying
$ e^{(a)} \cdot e^{(b)} = - \delta^{ab}$
and 
\begin{eqnarray}
-\sum_{a} (e^{(a)}    
\otimes e^{(a)})_{\mu \nu} &=& - ( e_\mu^{(1)} 
e_\nu^{(1)}+e_\mu^{(2)} e_\nu^{(2)} )  \equiv e_{\mu \nu} \;, 
\\
\sum_{a} (e^{(a)}    
\wedge e^{(a)})_{\mu \nu} &=&  e_\mu^{(1)} 
e_\nu^{(2)}- e_\mu^{(2)} e_\nu^{(1)}  \equiv  \epsilon_{\mu \nu}   \;.
\end{eqnarray}
In particular, one can choose
\begin{eqnarray}
e^{\mu\nu}&=&\eta^{\mu \nu}-\frac{(n\cdot k)}{D}
(n^{\mu}k^{\nu}  +n^{\nu}k^{\mu})    +\frac{k^2}{D} n^{\mu}n^{\nu}
+\frac{n^2}{D} k^{\mu}k^{\nu}\;,
\\
\epsilon^{\mu \nu}&=& \frac{1}{\sqrt{D}} \epsilon^{\mu 
\alpha \rho \nu}n_{\alpha}k_{\rho}\;,
\end{eqnarray}
with $D=(n\cdot k)^2-n^2k^2$.
With these elements the photon propagator can be written as
\begin{equation}\label{PROPAGATOR}
\Delta_{ \mu \nu }(k) = -\sum_{\lambda=\pm 1 } 
\frac{P^{(\lambda)}_{\mu \nu}(k)}{ k^2+2g\lambda (k\cdot n)^2\sqrt{D} }\;,
\end{equation}
where $P^{(\lambda)}_{\mu \nu}=\frac{1}{2}(e_{\mu\nu} 
+i\lambda \epsilon_{\mu \nu} )$ is an orthogonal projector.
%-------------------------------------------------------------------------
\section{The fermion self-energy}
%------------------------------------------------------------------------
We compute the fermion self-energy 
with the modifications 
introduced only via the Lorentz violating photon propagator \eqref{PROPAGATOR}.
The one loop-order approximation to the fermion self-energy is 
\begin{eqnarray}\label{SELF-ENERGY}
\Sigma_2(p)&=&ie^{2}   \int \frac{d^{4}k}
{(2\pi )^{4}}\gamma ^{\mu
}\left( \frac{  \sla{p}-\sla{k}
+m}{(p-k)^{2}-m^{2}}\right)  \gamma ^{\nu }
{  \Delta}_{\mu \nu }(k) \;,
\end{eqnarray}
which can be decomposed into a
$CPT$ even part
\begin{eqnarray}\label{even}
\Sigma_2^{(+)}(p)=-\frac{ie^{2}  }{2}   \sum_{\lambda } \int \frac{d^{4}k}
{(2\pi )^{4}}\gamma ^{\mu
}\left( \frac{  \sla{p}-\sla{k}
+m}{(p-k)^{2}-m^{2}}\right)
\frac{\gamma ^{\nu } e_{\mu\nu} }{ k^2+2g\lambda (k\cdot n)^2\sqrt{D} }\;,
\end{eqnarray}
and a $CPT$ odd part 
\begin{eqnarray}\label{odd}
 \Sigma_2^{(-)}(p)=-\frac{ie^{2}  }{2} \sum_{\lambda } \int \frac{d^{4}k}
{(2\pi )^{4}}\gamma ^{\mu
}\left( \frac{  \sla{p}-\sla{k}
+m}{(p-k)^{2}-m^{2}}\right)
\frac{  \gamma ^{\nu }i \lambda \epsilon_{\mu \nu}}{ k^2+2g\lambda (k\cdot n)^2\sqrt{D} } \;.
\end{eqnarray}
Next we expand in powers of external momenta obtaining
\begin{eqnarray}\label{expansion}
\Sigma_2(p)&=&\Sigma_2(0)+ p_{\alpha}
 \left( \frac{ \partial  \Sigma_2(p)}{\partial p_{\alpha}}\right)_{p=0} + \Sigma_g   \;,
\end{eqnarray}
where $\Sigma_g$ are convergent terms 
in the limit $g\to 0$ depending on quadratic and higher powers of $p$.

In order to compute the corrections our strategy will be 
i) perform a Wick rotation and
extend analytically any four vector to the Euclidean
$x_E=(ix_{0},\vec x)$, and ii) use dimensional regularization in spherical coordinates for the
divergent integrals.
To begin, we are interested on the first two even contributions in 
 Eqs. \eqref{even} and \eqref{expansion}, which are
\begin{eqnarray}
\Sigma_2^{(+)}(0)&=&-  \frac{ie^{2}  }{2} m   \sum_{\lambda } \int \frac{d^{4}k}
{(2\pi )^{4}}      \frac{1}{(k^{2}-m^{2})}    \frac{  \gamma ^{\mu }e_{\mu\nu} \gamma ^{\nu } }{k^2+2g\lambda (k\cdot n)^2\sqrt{D}   }  \;,
\\
 \frac{\partial  \Sigma_2^{(+)}(0)}{\partial p_{\alpha}}  &=&-\frac{ie^{2}  }{2} \sum_{\lambda }   \int \frac{d^{4}k}
{(2\pi )^{4}}   \left[    \frac{1}{(k^{2}-m^{2})}    -
 \frac{2k_ {\alpha}^2}{(k^{2}-m^{2})^2}  
  \right]  \\ &\times&  \frac{    \gamma^{\mu}\gamma^{\alpha}\gamma^{\nu} 
  e_{\mu \nu}}{k^2+2g\lambda (k\cdot n)^2\sqrt{D}   } \nonumber  \;.
\end{eqnarray}
Applying our strategy leads to
\begin{eqnarray}
\Sigma_2^{(+)}(0)&=& e^{2} m   \sum_{\lambda } \int \frac{d^{4}k_E}
{(2\pi )^{4}}   \frac{1}{(k_E^{2}+m^{2}) (k_E^2-2g \lambda (k_E\cdot n_E)^2 \sqrt{D_E})}\nonumber \;,
\\
\frac{\partial  \Sigma_2^{(+)}(0)}{\partial p_{\alpha}}  &=&-
\frac{e^{2}}{2}  (n_{\nu} n^{\alpha}-\frac{n_E^2}{2}\eta^{\alpha}_{\nu}    ) 
 \gamma^{\nu}   \sum_{\lambda } \int \frac{d^{4}k_E} 
{(2\pi )^{4}}  \left[    \frac{1}{(k_E^{2}+m^{2}) }
 -\frac{k_ E^2}{2(k_E^{2}+m^{2})^2}  \right] \nonumber  \\ 
 &\times& \frac{k_E^2}{D_E}\frac{1}{(k_E^2-2g \lambda (k_E\cdot n_E)^2 \sqrt{D_E})}\;,
\end{eqnarray}
where we have used $\gamma ^{\mu }e_{\mu\nu} \gamma ^{\nu } =2$
and $D_E=(n_E\cdot k_E)^2-k_E^2n_E^2$.
The calculation produces
\begin{eqnarray}\label{mass}
\Sigma_2^{(+)}(0)&=& \frac{e^2m}{8\pi^2} \left(1-\ln \left(\frac{g^2m^2(n_E^2)^3}{16}\right)\right)\;,
\\
 p_{\alpha} \frac{\partial  \Sigma_2^{(+)}(0)}{\partial p_{\alpha}}&=&-
  \frac{e^{2}}{16\pi^2}  \left(\frac{1}{2}\sla{p}-\frac{\sla{n} (n\cdot p)}{n_E^2}    
\right)  \left(1+ \ln \left(\frac{g^2m^2(n_E^2)^3}{16}\right)    \right) \nonumber \;.
\end{eqnarray}
Let us emphasize that the renormalization in the even sector
 involves small
corrections without any possible fine-tuning. 
Also, the radiative corrections to the 
mass and wave function
are finite and have the usual
 logarithmic divergence in the limit $g\to 0$.

Now we compute the lower dimensional operator
 $\bar \psi \sla{n} \gamma_5 \psi $ which arises 
in the radiatively correction to the odd sector.
 According to Eqs. \eqref{odd} and
 \eqref{expansion} it comes from
\begin{eqnarray}
 \Sigma_2^{(-)}(0)=-2ge^{2} (\epsilon_{\mu \alpha \beta \nu} n^{\alpha}
 \gamma^{\mu}   \gamma^{\sigma} \gamma^{\nu})   \int \frac{d^{4}k }{(2\pi )^{4}}
\frac{k^{\beta}k_{\sigma} }{(k^2-m^2)}
\frac{(n\cdot k)^2}{((k^2)^2-4g^2(k\cdot n)^4D)}    \;.
\end{eqnarray}
We can extract the correction
from the most general form of the above integral 
$
 F\, \delta^{\beta}_{\sigma}+R \,n ^{\beta} n_{\sigma}\;
$
and considering $\epsilon_{\mu \alpha \beta \nu} n^{\alpha}
\gamma^{\mu} \gamma^{\beta}\gamma^{\nu} =3!i\sla{n}\gamma^5$
which requires to find 
\begin{eqnarray}\label{F}
 F=-\frac{2ge^{2} }{3n^2} \int \frac{d^{4}k }{(2\pi )^{4}}
\frac{ D (n\cdot k)^2}{(k^2-m^2)
((k^2)^2-4g^2(k\cdot n)^4D)}\;.\nonumber 
\end{eqnarray}
For this divergent element we have
 in $d$ dimensions
\begin{eqnarray}
F&=&\frac{2ig e^{2} n_{E}^2}{(d-1)} \mu^{4-d}\int   \frac{ d\Omega }
{(2\pi )^{d}}        \sin^2\theta   \cos^2\theta   
\int_{0}^{\infty} \frac{ d|k_E| \, |k_E|^{d-1} M^2 }{(|k_E|^2+m^2)
\left(|k_E|^2+M^2   \right)}\;,  \nonumber  
\end{eqnarray}
where $M^2=\frac{1}{4g^2 n_E^6 \sin^2 \theta \cos^4 \theta}$,
and $\theta$ is the angle
 between $n_E$ and $k_E$ and  $\left|k_E \right| =\sqrt{k_E^2}$.
Next considering the solid angle element in $d$ dimensions,
$\int d\Omega=  \frac{ 2\pi^{\frac{d-1}{2}}}
{\Gamma(\frac{d-1}{2})}\int_0^\pi d\theta
 (\sin\theta)^{d-2}$,
and using the approximation $M^2\gg m^2$ 
with
 $d=4-\varepsilon$
the dominant contribution is 
\begin{eqnarray}\label{final}
F &=&-\frac{ie^2 }{\varepsilon \pi^2 }
\left(\frac{1}{{ 24g(n^2)^2  }} + \frac{g m^2 n^2}{96}\right) 
 +\frac{i e^2}{1152 g  \pi^2(n^2)^2} \Big[- 24 \ln(
   g^2  \pi \mu^2 (n^2)^3))  \nonumber 
\\ &&-160  
+24 \gamma_E  + 8\ln(64) \Big]\;.
\end{eqnarray}
At this level in the radiative corrections the presence of the
 high scale $g$ is to fine-tune the parameters 
in the $CPT$ odd sector of the theory and produce small
and finite 
 corrections in the $CPT$ even sector \eqref{mass}.
To deal with the large Lorentz corrections 
in the odd sector  
we will take a step further in the renormalization program. 
In the presence of Lorentz-$CPT$ violation a renormalization program 
is far from being trivial, however, a systematic method exist in order to 
generalize the LSZ reduction formalism and pole extraction \cite{Potting, lehnert}.
We consider some elements of the method in order to 
generalize the renormalization conditions and the expression for the two-point function $\Sigma_R$.

Starting from the Lagrangian (\ref{Lag}), we renormalize the 
electron wave function and mass with $m=Z_mm_R$, $\psi=\sqrt{Z_2} \psi_R$ 
initially present in the Lagrangian. Replacing 
$Z_2=1+\delta_2$, $Z_m=1+\delta_m$ in the
Lagrangian, the renormalized Green function 
can be written as
\begin{eqnarray}\label{RGreen}
G^{(2)-1}_R= \sla{p}-m_R  +\delta_2\sla{p}-(\delta_2+\delta_m)m_R
+\Sigma_2   \;.
\end{eqnarray}
Our calculation shows that 
 \begin{eqnarray}
\Sigma_2&=&A \sla{p}+Bm_R+\sum_{i=1}^{4} f_iM_i    \;,
\end{eqnarray}
where the coefficients $A, B$ and $f_i$ can
depend on the scalars $p^2, (n\cdot p)$ and we have defined
$M_1=\sla{n},  M_2=\sla{n}\gamma^5, M_3=\sla{p}\gamma^5,
M_4=[\sla{n},\sla{p}] \gamma^5$.
Let us consider the ansatz $\bar P=\sla{p}-\bar m+ \sum_{i=1}^{4} \bar x_i M_i$,
where $\bar m=m_R+m_n(n\cdot p)$ and the coefficients $\bar x_i$ are independent 
of the previous scalars but depend linearly on the perturbative parameter $\alpha=\frac{e^2}{4 \pi}$, see \cite{lehnert}.
Replacing $\bar P$ in \eqref{RGreen}, we have 
\begin{eqnarray}
G^{(2)-1}_R(\bar P) = \bar P+\Sigma_R(\bar P)   \;,
\end{eqnarray}
and
\begin{eqnarray}
\Sigma_R(\bar P)&=&  m_n(n \cdot p)  - \sum_{i=1}^{4} \bar x M_i + \delta_2\left(\bar P+m_n(n \cdot p)- 
\sum_{i=1}^{4} \bar x M_i   \right)\nonumber \\&& -\delta_mm_R+\Sigma_2(\bar P) \;.
\end{eqnarray}
Demanding the Green function $G^{(2)}_R$ to have a 
 a pole at $\bar P=0$ and residue $i$
we obtain two renormalization conditions
\begin{eqnarray}
\delta _mm_R&=& m_n(n\cdot p)-\sum\bar x_i M_i +\delta_2\left(m_n  (n\cdot p)- \sum\bar x_i M_i  
\right)  +   \Sigma_2(0) \nonumber \;,
\\
\delta_2&=&-\frac{d \Sigma_2(0)}{d\bar P}   \;.
\end{eqnarray}
Replacing in the expression for $\Sigma_{R}(\bar{P})$ leads to
\begin{eqnarray}\label{contr}
\Sigma_{R}(\bar{P})=\Sigma_2(\bar{P})-\Sigma_2(0) -\frac{1}{2}\left\{
\bar{P} , \,\frac{d \Sigma_2(0)  }{d \bar{P}}  \right\}\;,
\end{eqnarray}
which due to the non commutativity of 
$\bar P$ and $\frac{\partial \Sigma_2(0)  }{\partial \bar{P}}$
 presents an order ambiguity. However, at lowest order in perturbation theory
 the 
 contribution from the derivative part 
comes from $m_R \frac{d \Sigma_2(0)  }{d \bar{P}}$
which is free of the order ambiguity.

Considering the modified dispersion relation satisfied by $\bar P$ and focusing 
on the odd contributions we have
\begin{eqnarray}
 &&(\Sigma_2^{(-)})(\bar P)-( \Sigma_2^{(-)})(0)=-
\left(\epsilon_{\mu \alpha \beta \nu}
 \gamma^{\mu}\gamma^{\beta}\gamma^{\nu} 
 n^{\alpha} \right)    \frac{4ign^2_E
 e^{2} \mu^{4-d}}{(2\pi )^{d}} \int_0^1 dx   \int _0^{1-x}   dy\nonumber \\
 &&\times \int 
d|k_E|  |k_E| ^{d+1} d\Omega    M^2 \cos^2 \theta \left( \frac{ 1 }
{(k_E^2+Q_1)^3}-\frac{ 1 }{( k_E^2+Q_2)^3}  \right)  \;,
\end{eqnarray}
where we have considered the Feynman parametrization and 
defined $Q_1=m^2 x+M^2y$, $Q_2=m^2 x^2+M^2y $ and dropped the label 
$R$ for the physical mass.  The scalar part above is a finite term
\begin{eqnarray}
\frac{ign^2_E e^{2} }{8 \pi ^{4}} \int_0^1dx  
\int_0^{1-x} dy    \int d\Omega
 M^2  \cos^2 \theta   \ln \left(\frac{Q_2}{Q_1} \right)\;.
\end{eqnarray}
With the approximation
\begin{eqnarray}
  \ln \left(\frac{Q_2}{Q_1} \right)&=&\ln \left(1-\frac{   
 m^2x(1-x)}{M^2y+m^2x}\right) \approx - \frac{   m^2x(1-x)}{M^2y+m^2x}\;,
\end{eqnarray}
and integrating in $x$ and $y$ we find
\begin{eqnarray}\label{domcont}
 (\Sigma_2^{(-)})(\bar P)-(\Sigma_2^{(-)})(0)=\frac{gm^2 \alpha
 n_E^2\sla{n}_E\gamma_5}{8\pi} \left(3-12\ln(2)+2\ln(4g^2m^2n_E^6)\right)\;,
\end{eqnarray}
where $\alpha=\frac{e^2}{4 \pi}$.
From \eqref{contr} and after some algebra
 the derivative contribution is  found to be 
\begin{eqnarray}\label{dercontr}
 m\frac{d \Sigma_2^{(-)}(0) }{d \bar P} &=&\left(\epsilon_{\mu \alpha \beta \nu}
 \gamma^{\mu}\gamma^{\beta}\gamma^{\nu} n^{\alpha} \right) 
  \frac{-gie^{2} m^2  n_E^2\alpha}{72\pi}\ln(2gm |n_E|^3)
 \;.
\end{eqnarray}
Finally, the dominant contribution from \eqref{domcont} and \eqref{dercontr} is
\begin{equation}
 \Sigma^{(-)}_{R}(0)=\frac{7gm^2 n_E^2  \alpha  \ln(2gm |n_E|^3) }{12\pi} \sla{n}\gamma_5\;.
\end{equation}
By considering the bound  $ \Sigma^{(-)}_{R}(0) <10^ {-31}$ GeV, coming from
a torsion pendulum experiment
\cite{Table,torsion}, we find the bound 
\begin{eqnarray}
\xi<6 \times 10^{-3}\;.
\end{eqnarray}
%%%%%%%%%%%%%%%%%%%%%%%%%%%%%%%%%%%%
\section{Conclusions}
%%%%%%%%%%%%%%%%%%%%%%%%%%%%%%%%%%%%
Effective field theory provides a very powerful tool in order to
check for consistent Lorentz symmetry violation at low energies.
This is specially true for effective theories 
with
higher-order operators where operators
generated via radiative corrections are
unprotected against 
fine-tuning. However, in the Myers and Pospelov model we have shown that
 the same symmetries 
that allows an operator to be induced 
 will also dictate the size of the
correction.

We have considered the even and odd $CPT$
parts coming from modifications in the photon propagator.
 We have shown that the radiative corrections to the even
 $CPT$ sector are given by small contributions
  to the usual parameters of the standard model couplings.
On the contrary, in the odd $CPT$ sector we have found large Lorentz violations
 in the induced axial operator of mass dimension-3.
For the calculation we have used
 dimensional regularization in order to preserve unitarity 
 and considered a general background which incorporates the 
effects of higher-order time derivatives.
The large Lorentz violation has been shown to be controlled by defining 
 the on-shell mass subtraction for the fermion, leading to
acceptable small Lorentz violating radiative corrections.
We leave for future work the full renormalization of this model by 
taking into account all the Feynman diagrams.
%------------------------------------------------------------------------
\section*{Acknowledgments}
%------------------------------------------------------------------------
C.M.R. thanks Markos Maniatis for  
helpful discussions and acknowledges 
support from the projects FONDECYT REGULAR \# 1140781
and DIUBB \# 141709 4/R.
%%%%%%%%%%%%%%%%%%%%%%%%%%%%%%%%

\end{document}